\documentclass{article}

\usepackage[utf8]{inputenc}
\usepackage{amsmath}
\usepackage{amssymb}
\usepackage{authblk}
\usepackage{biblatex}
\usepackage{verbatim}
\usepackage{subcaption}
\usepackage{graphicx}
\usepackage{placeins}

\title{Decentralized Distributed Proximal Policy Optimization (DD-PPO) for High Performance Computing Scheduling on Multi-User Systems}
\author[1]{Sgambati, Matthew}
\affil[1]{Idaho National Laboratory}
\author[2]{Vakanski, Aleksandar}
\affil[2]{University of Idaho}
\author[1]{Anderson, Matthew}
\date{March 2025}

\begin{document}

\maketitle

\section*{Abstract}
\sloppy
Resource allocation in High Performance Computing (HPC) environments presents a complex
and multifaceted challenge for job scheduling algorithms. Beyond the efficient allocation
of system resources, schedulers must account for and optimize multiple performance metrics,
including job wait time and system utilization. While traditional rule-based scheduling
algorithms dominate the current deployments of
HPC systems, the increasing heterogeneity and
scale of those systems is expected to challenge
the efficiency and flexibility of those algorithms in minimizing
job wait time and maximizing utilization. Recent research efforts have focused
on leveraging advancements in
Reinforcement Learning (RL) to develop more adaptable and intelligent scheduling strategies.
Recent RL-based scheduling approaches have explored a range of algorithms, from Deep
Q-Networks (DQN) to Proximal Policy Optimization (PPO), and more recently, hybrid methods that integrate Graph Neural
Networks (GNNs) with RL techniques. However, a common limitation across these methods is their reliance on relatively
small datasets, and these methods face scalability issues
when using large datasets.
This study introduces a novel RL-based scheduler utilizing the Decentralized Distributed Proximal Policy Optimization
(DD-PPO) algorithm, which supports large-scale distributed training across multiple workers without requiring parameter
synchronization at every step. By eliminating reliance on centralized updates to a shared policy, the DD-PPO scheduler
enhances scalability, training efficiency, and sample utilization. The validation dataset
leveraged over 11.5 million real HPC job traces for
comparing DD-PPO performance between traditional and advanced
scheduling approaches, and the experimental results demonstrate improved scheduling performance in comparison to both
rule-based schedulers and existing RL-based scheduling algorithms.

\section{Introduction}
Resource management is a critical task across various systems, from virtual machines~\cite{vm_placement_abohamama}
to job scheduling~\cite{job_scheduling_zhan}. These systems typically rely on traditional rule-based algorithms,
such as First Come, First Served (FCFS) and Shortest Job First (SJF)~\cite{pinedo2012scheduling}, to efficiently
allocate resources. In High Performance Computing (HPC), job schedulers like PBS~\cite{pbs} and SLURM~\cite{slurm} utilize
related rule-based approaches to optimize scheduling decisions based on metrics such as
job submission time, estimated run time, and requested resources.
More advanced algorithms, such as UNICEF~\cite{unicef} and F1~\cite{f1}, build upon
rule-based approaches by incorporating additional job attributes to improve scheduling
quality. For example, techniques like backfilling and reserving resources for large jobs
not only refine scheduling decisions but also enhance overall resource utilization.
However, as HPC systems continue to
grow in scale and heterogeneity ~\cite{hpc_scale_problem1,hpc_scale_problem2}, traditional rule-based methods have less
flexibility to optimize system utilization and decrease job wait time for multi-user systems 
with a broad mix of resource and walltime requests.

To overcome this, recent scheduling efforts have
centered on leveraging Reinforcement Learning (RL) approaches which have famously shown success
in Chess and Go and produce models that are optimal even when the environment changes.  Some of these
RL efforts for scheduling have included Proximal Policy Optimization (PPO)~\cite{ppo},
Deep Q Networks (DQN)~\cite{dqn}, and Soft Actor-Critic (SAC)~\cite{sac}. Such algorithms are particularly well-suited
for HPC job scheduling due to their ability to interact dynamically with the environment to continuously adapt their
policies based on the feedback and rewards they receive. This capacity allows RL methods to learn and respond to the
complexities of the ever-changing HPC environments that they are running in. Adjusting to
a changing environment is difficult both for 
unsupervised and supervised machine learning approaches in general as well as
rule-based approaches.
But in order to produce optimal results in
the presence of environmental changes, RL scheduling algorithms need to be trained on large volumes
of data, which in turn demands significant computing capabilities~\cite{large_data_ai}. 
This is compounded by a limitation in
scalability that current RL
schedulers struggle with due to centralized policy updates, limiting training efficiency and
performance on large-scale datasets.

This study introduces an RL-based scheduler utilizing Decentralized Distributed Proximal Policy
Optimization (DD-PPO)~\cite{ddppo}, which supports large-scale distributed training across multiple workers without
requiring parameter synchronization at every step. In our study, the DD-PPO algorithm is implemented in the Ray~\cite{ray}
framework. Experimental validation using a large real-world dataset containing more than 11.5 million job traces collected over
six years demonstrates better performance in comparison to both rule-based schedulers and existing RL-based
scheduling algorithms. By eliminating reliance on centralized updates to a shared
policy, the DD-PPO scheduler enhances scalability, efficiency, and sample utilization. This scheduler is a
general-purpose HPC scheduling system that utilizes a large training dataset to learn intricate system-specific and
workload characteristics across diverse HPC environments. By capturing these patterns, it enhances adaptability to
previously unseen features, potentially offering greater flexibility compared to specialized, customized scheduling methods.

DD-PPO offers several distinct advantages over PPO, the most common approach in recent RL schedulers. First,
DD-PPO is highly scalable because it is specifically designed for distributed environments, allowing seamless
expansion across multiple machines and GPUs. Additionally, by leveraging distributed training, DD-PPO improves
sample efficiency through the parallel collection and processing of large batches of samples, which enhances
its learning capability. Another key benefit is faster convergence; parallel data processing accelerates the
learning process and reduces overall training time. Moreover, the increased computational power and sampling
efficiency of DD-PPO make it more effective at handling complex environments and tackling challenging tasks.
Finally, DD-PPO demonstrates strong robustness: utilizing diverse data gathered from multiple workers helps
minimize overfitting and instability during training, thereby improving its generalization and adaptability to
unseen scenarios.

The structure of this work is as follows. Section~\ref{sec:related_work} presents a review of
related work and existing RL-based scheduling algorithms. An overview of the foundational concepts involved in
implementing RL algorithms for HPC job scheduling is covered in section~\ref{sec:preliminaries}. The methodology
of the proposed algorithm, datasets, and scheduling framework used is provided in section~\ref{sec:methodology}.
Section~\ref{sec:results} provides the results of the experimental validation and performance comparison against
rule-based and RL-based algorithms. Section~\ref{sec:conclusions} provides conclusions of the scheduling strategy
based on DD-PPO and discusses avenues for future work.

\section{Related Work}
\label{sec:related_work}
Despite recent advancements in RL-based scheduling, its application in HPC environments
still poses various challenges. Many of the advancements have leveraged RL-based scheduling
techniques that are not designed for or geared towards HPC batch job scheduling. Early examples
of deep reinforcement learning (DRL) schedulers include DeepRM~\cite{deeprm} and Decima~\cite{decima}.
DeepRM is a simple multi-resource cluster scheduler that uses a standard policy gradient
(PG) RL algorithm trained using synthetic datasets of job traces. Decima utilizes RL to
optimize the allocation of data  processing jobs, with the jobs being composed of dependent tasks and structured
into a directed acyclic graph. In this approach most HPC jobs consist of a single large task that runs from start
to finish, making them rigid and non-composable. As a result, Decima is not well-suited for dynamic and heterogeneous
HPC environments, as it depends on job malleability for effective scheduling.

Zhang et al.~\cite{rlscheduler} proposed RLscheduler, a general DRL-based scheduling
model that is trained using a single system log and later applied to different systems
with varying characteristics, such as system size and workload patterns. RLScheduler is
trained on individual datasets that are comprised of either synthetic or real workloads.
Fan et al.~\cite{dras} introduced an automated scheduling agent
DRAS, which leverages DRL. DRAS incorporates a hierarchical neural network with specialized
HPC scheduling features like resource reservation and backfilling, allowing it to dynamically
adjust policies based on workload changes. DRAS comprises four distinct customized
agents based on Deep Q-learning, Policy Gradient, Advantage Actor-Critic, and PPO, with PPO providing
the best results. It is trained on both synthetic data and real workloads. However, despite having
access to over 2.7 million real job traces, the training dataset consists primarily of synthetic traces,
which make up about 90\% of the data.
IRLS scheduler~\cite{irls} extends the DRAS approach by incorporating
additional identity features within the state definition to enhance job runtime prediction.
Unlike DRAS, IRLS was trained using job traces from the smaller SDSC-SP2-1998 dataset.
Wang et al.~\cite{deep_back} developed RLSchert, a job scheduler built upon PPO. Its key
enhancement lies in its ability to estimate the remaining runtime of jobs by utilizing a
dynamic job runtime predictor, which is trained on features extracted from HPC jobs and
uses imitation learning to create the optimal policy for selecting or terminating jobs
based on the system. RLSchert is trained on a small real workload set of job traces based entirely on a
single HPC software called VASP~\cite{vasp}. VASP is a computational software used for atomic-scale materials
modeling. SchedInspector by Zhang et al.~\cite{schedinspector} incorporates runtime factors into multiple
batch job scheduling policies to enhance job execution performance. It utilizes PPO and a key advantage
is its ability to automatically adapt to and improve existing scheduling policies without modifying them.
One key drawback is that while it improves job execution performance by integrating runtime factors into
scheduling decisions, it may leave resources idle when rejecting scheduling choices, which can impact
system utilization. It was trained on both synthetic and real workload job traces.

These prior studies demonstrate significant progress in RL-based scheduling and advantageous performance
compared to rule-based algorithms. However, apart from the DRAS approach, most rely on small datasets and
utilize PPO. Additionally, schedulers face several limitations that hinder their adoption in real-world applications.
They require extensive computational resources, suffer from sample inefficiency, and can struggle to generalize across
diverse scheduling environments. RL models also face challenges in balancing exploration and exploitation, handling
complex constraints, and maintaining interpretability, making them less reliable than traditional scheduling methods.
For these reasons, industry adoption remains low due to integration challenges, risk aversion, and reliance on established
deterministic scheduling techniques. To make RL schedulers viable, advancements are needed in sample efficiency,
generalization, hybrid approaches, interpretability, and constraint handling.

\section{Preliminaries}
\label{sec:preliminaries}
This section outlines the foundational concepts involved in implementing RL algorithms. It covers
elements such as state and action space design, reward formulation, environment modeling, and integration with scheduling
frameworks, which are essential for developing effective RL-based schedulers capable of learning and adapting to the dynamic
and resource-intensive nature of HPC workloads.

\subsection{RL Overview}
RL is a computational framework for acquiring knowledge through interaction with an environment. In this framework,
an agent explores an environment, makes decisions, receives feedback in the form of rewards or penalties, and refines
its strategy to optimize cumulative rewards over time~\cite{rl}. The environment that the agent explores is modeled by a
Markov decision process (MDP), which is a mathematical framework for modeling decision-making in stochastic
environments where outcomes are partly random and partly under the control of an agent. An MDP is defined by the tuple
$(S, A, P, R, \gamma)$, where $S$ represents the set of states, $A$ denotes the set of possible actions, $P(s'|s,a)$ is
the transition probability from state $s$ to state $s'$ given action $a$, $R(s,a)$ is the reward function, and $\gamma \in
[0,1]$ is the discount factor that determines the weight of future rewards. The agent seeks to learn an optimal policy
$\pi(a|s)$ that maximizes the expected cumulative reward over time.

\begin{figure}[h]
    \centering
    \includegraphics[width=0.65\linewidth]{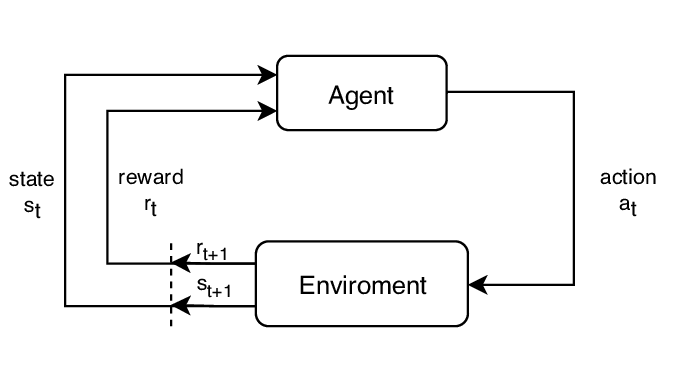}
    \caption{Diagram showing the general framework of reinforcement learning.}
    \label{fig:basic_rl_diagram}
\end{figure}

Figure~\ref{fig:basic_rl_diagram} shows the general framework for RL. When using MDPs, learning agents engage with a changing
environment at distinct time intervals. At each timestep $t$, the agent observes the current state $S_t$ and
selects an action $A_t$. This action results in a transition of the environment from state $S_t$ to $S_{t+1}$, while the
agent receives a reward $R_{t+1}$ as feedback. Typically, the agent has no prior knowledge of the environment's dynamics
or reward structure and must gradually learn them through interaction during training. The objective of reinforcement
learning is to maximize the expected cumulative discounted reward:

\begin{equation}
G_t = \sum_{k=0}^{\infty} \gamma^k R_{t+k},
\end{equation}

\noindent where $\gamma$ is the discount factor controlling the importance of future rewards. The agent follows a policy
$\pi(a|s)$, which defines the probability distribution of selecting specific actions in given states. 

There are two primary approaches to RL algorithms, value-based and policy-based. Value-based RL algorithms estimate the
expected return for given states or actions. A fundamental approach is Q-learning, which iteratively updates the
Q-value to refine action selection~\cite{watkins1992q}:

\begin{equation}
Q(s,a) \leftarrow Q(s,a) + \alpha [R + \gamma \max_{a'} Q(s',a') - Q(s,a)].
\end{equation}

To handle high-dimensional state spaces, DQN extends Q-learning by utilizing neural networks to
approximate Q-values, enabling effective decision-making in complex environments~\cite{mnih2015human}.

In contrast, policy-based algorithms focus on directly optimizing the policy function $\pi_{\theta}(a|s)$,
parameterized by $\theta$. One widely used method is REINFORCE~\cite{williams1992reinforce}, which updates
policy parameters through gradient ascent~\cite{williams1992simple}:

\begin{equation}
\theta \leftarrow \theta + \alpha \nabla_{\theta} J(\theta),
\end{equation}

\noindent where the objective function $J(\theta)$ is given by:

\begin{equation}
J(\theta) = \mathbb{E} \left[ G_t \log \pi_{\theta}(a_t | s_t) \right].
\end{equation}

\subsection{Actor-Critic Method}
In RL, one primary approach is through value-based methods. These methods focus on minimizing
a loss function by solely approximating a value function. Typically, they use an objective
function based on the Bellman equation and operate in an off-policy manner, meaning they can
leverage data collected at any point during training.

An alternative approach is policy-gradient methods, which aim to maximize the performance of
a parameterized policy that selects actions without consulting a value function. These methods
are on-policy, so they can only utilize data from the most recent version of the policy during
optimization. Policy-gradient methods offer several advantages:

\begin{itemize}
    \item They perform better in partially observed environments because arbitrary action
    probabilities can be learned.
    \item They have the potential to converge to a deterministic policy.
    \item The action probabilities adapt continuously as a function of the learned parameters.
\end{itemize}

Moreover, policy-gradient methods tend to be more stable since they directly optimize the policy
rather than relying on a separate value function to derive actions. However, value-based methods
remain more sample efficient, as they can reuse data collected at any point during training.

Policy-gradient methods and value-based methods are not mutually exclusive. Actor-Critic methods are
another approach that integrates elements of both value-based and policy-based approaches, enhancing
stability and performance. They are methods that learn both a policy and value function through two
separate structures that interact with each other. The policy structure is typically known as the actor
since it selects actions to take and the value function structure is known as the critic since it criticizes
actions made by the actor. The actor updates the policy, while the critic evaluates it through a value
function~\cite{konda2000actor}:

\begin{equation}
\delta_t = R_t + \gamma V(s_{t+1}) - V(s_t).
\end{equation}

\noindent The policy is then refined in the direction of the advantage:

\begin{equation}
\theta \leftarrow \theta + \alpha \delta_t \nabla_{\theta} \log \pi_{\theta}(a_t | s_t).
\end{equation}

By merging the benefits of Q-learning with policy optimization, Actor-Critic methods provide a balanced RL
framework that leads to more robust and efficient policy updates. Furthermore, the Generalized Advantage
Estimation (GAE)~\cite{gae} technique can be employed to bolster the stability of the actor's updates. GAE
quantifies how much better an action is compared to the average action at a given state by incorporating a
$\lambda$ parameter to control the trade-off between bias and variance.

\subsection{Proximal Policy Optimization}
PPO is an on-policy Actor-Critic RL algorithm that alternates between sampling data from
the environment and optimizing a clipped surrogate objective via stochastic gradient descent. The clipping
mechanism is crucial as it prevents excessively large policy updates, thereby enhancing training stability
compared to other RL approaches. While PPO can employ a penalty term to restrict drastic policy changes,
studies have shown that clipping is both simpler and more effective~\cite{ddppo}. The PPO objective function
is defined as:

\begin{equation}
L(\theta) = \mathbb{E}_t \left[ \min \left( r_t(\theta) A_t,\; \operatorname{clip}(r_t(\theta), 1-\epsilon, 1+\epsilon) A_t \right) \right],
\label{eq:ppo}
\end{equation}

\noindent where the probability ratio of the new policy, denoted as $r_t(\theta)$, is defined as
$\frac{\pi_{\theta}(a_t | s_t)}{\pi_{\theta_{\text{old}}}(a_t | s_t)}$, which measures how the
updated policy $\pi_{\theta}$ compares to the previous policy $\pi_{\theta_{\text{old}}}$. The advantage
estimate, $A_t$, quantifies the benefit of taking a specific action relative to the expected outcome,
thereby guiding the policy update. To ensure stability, the policy adjustment is constrained using the
clipping function $\text{clip}(r_t(\theta), 1 - \epsilon, 1 + \epsilon)$, where the clipping threshold
\(\epsilon\) is typically set to a small value (e.g., 0.2) to limit drastic changes between updates.

\section{Methodology}
\label{sec:methodology}
This section outlines our proposed method for HPC scheduling, highlighting its design, implementation, and
benefits. We begin by providing an overview of the DD-PPO algorithm, which forms the basis of our approach,
and then describe our hyperparameter tuning process.

\subsection{Decentralized Distributed Proximal Policy Optimization}
\label{sec:ddppo}
DD-PPO enhances the standard PPO algorithm by enabling efficient training across multiple machines and GPUs.
It achieves this by decentralizing the learning process—each worker independently collects experiences and
computes gradients, which are then periodically synchronized without relying on a centralized parameter
server. DD-PPO offers three key features:
\begin{itemize}
    \item \textbf{Decentralization:} Each worker gathers experiences and computes gradients autonomously.
    \item \textbf{Synchronization:} Gradients are averaged across workers at regular intervals to maintain
    consistency and prevent stale updates.
    \item \textbf{Scalability:} Near-linear scaling makes it well-suited for computationally intensive tasks.
\end{itemize}

The objective function in DD-PPO follows the standard clipped surrogate loss from PPO (see Equation~\ref{eq:ppo}).
In addition, DD-PPO employs distributed gradient synchronization with the following update rule:

\begin{equation}
\theta \leftarrow \theta + \alpha \sum_{i=1}^{N} \nabla_{\theta} L_i(\theta),
\end{equation}

\noindent where \(N\) denotes the total number of distributed workers.

In this framework, each worker \(i\) collects a batch of experiences from the environment and computes its
local gradient \(\nabla_{\theta} L_i(\theta)\) based on its own loss function \(L_i(\theta)\). This local loss
is calculated from the clipped surrogate objective, as in the standard PPO algorithm. Once all workers have
computed their gradients, these gradients are aggregated (by summing) across all \(N\) workers. This summed
gradient represents a more robust estimate of the true gradient over the entire distributed dataset.

The parameter update is then performed by scaling this aggregated gradient by the learning rate \(\alpha\)
and applying it to the shared model parameters \(\theta\). This synchronization ensures that each update step
benefits from the diversity of experiences collected across multiple workers, reducing the variance of the
gradient estimates and leading to more stable learning. By combining contributions from all workers, the
algorithm effectively increases the batch size and improves convergence properties, which is essential for
large-scale RL tasks.

By eliminating bottlenecks in experience collection, DD-PPO facilitates efficient, large-scale reinforcement
learning while retaining the stability and performance improvements of PPO. Moreover, DD-PPO can be coupled
with Population-Based Training (PBT)~\cite{pbt} for hyperparameter optimization. PBT simultaneously trains a
population of models, periodically selecting top performers to share their hyperparameters while introducing
random modifications to explore new configurations, thus enabling adaptive learning schedules.

Despite its benefits, DD-PPO also presents some drawbacks compared to PPO, such as increased implementation
complexity, greater sensitivity to hyperparameters, and a higher risk of instability. These challenges can be
mitigated by leveraging existing frameworks, exploiting the algorithm’s scalability to run numerous
simulations concurrently, and integrating techniques like PBT and GAE during training.

\subsection{Proposed Framework}
\begin{figure}[ht]
    \centering
    \includegraphics[width=.98\linewidth]{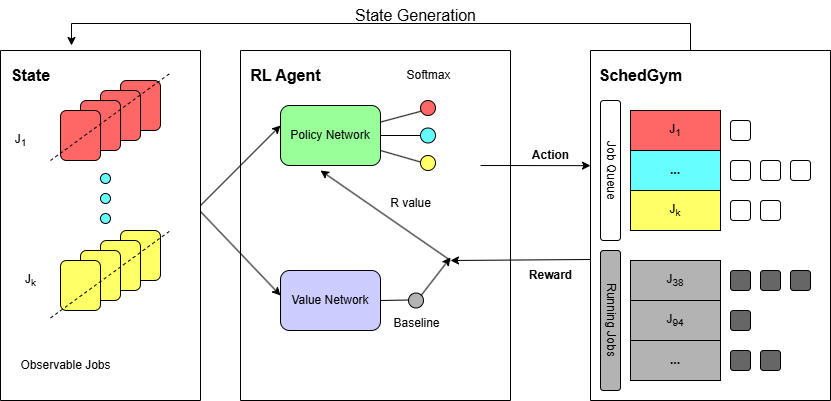}
    \caption{Overall architecture of our approach.
    Left box: The observable jobs are updated once an action is performed in the environment.
    Middle box: These jobs are sent as inputs into the RL agent policy and value networks.
    Right box: The agent next performs an action, which in turn updates the  state and the environment,
    which returns a reward to the agent based on the optimization goal.}
    \label{fig:hpc_rl_scheduler_overview}
\end{figure}

Our proposed framework leverages RL to develop adaptive scheduling policies for HPC batch jobs,
dynamically adjusting to varying workloads and optimization objectives. It takes job traces and
optimization goals as input, learning scheduling strategies autonomously. Figure~\ref{fig:hpc_rl_scheduler_overview}
shows the architecture and three main components: the Agent, the job scheduling Environment, and
the environment State. At each stage of the training process, the agent analyzes the current
state and selects an action. The state is derived from the environment, and the chosen action is
then applied, leading to the creation of a new state and the assignment of a reward. Over
multiple iterations, the agent refines its decision-making by learning from past actions and
their corresponding rewards.
The rewards are the feedback from the environment based on the action taken by the agent and are
used to guide the agent towards a better policy. The reward is calculated based on
the selected optimization goal. For this task, the goal is to minimize average waiting time,
average turnaround time, or average bounded slowdown, or to maximize resource utilization.

\begin{table}[b]
\centering
\resizebox{0.5\textwidth}{!}{%
\begin{tabular}{|c|c|c|}
\hline
\textbf{Name} & \textbf{Layers} & \textbf{Layer Size} \\ \hline
Policy        & 3               & 32, 16, 8           \\ \hline
Value         & 3               & 64, 32, 8           \\ \hline
\end{tabular}%
}
\caption{The network configurations of the policy and value networks.}
\label{tab:actor_critic}
\end{table}

The agent uses a policy network and a value network following the Actor-Critic model.
In our proposed approach, both the Actor and Critic are parameterized deep neural networks,
having the architectures shown in Table~\ref{tab:actor_critic}. The policy network is sequentially
applied to each waiting job. For every job, the network
computes and assigns a score, representing its priority value. These individual
scores collectively form a vector, which is then processed using a softmax function to generate a
probability distribution across all waiting jobs. As a result, when the jobs are reordered, their
corresponding probabilities are adjusted accordingly, ensuring a consistent ranking structure.
The value network receives a complete job sequence as input and produces a value representing the
expected reward for that sequence. These two networks are trained alongside each other. Once the
policy network finishes the scheduling decisions for a given job sequence, the corresponding
rewards are collected and then utilized to train the value network to enhance its ability to
estimate the expected reward for a given job sequence. The combination of these two networks
forms our Actor-Critic model.

In order to train the model effectively, a simulation environment for the RL agents to train
in needs to be utilized. We used an OpenAI Gym toolkit~\cite{openai_gym} based environment
called SchedGym. It takes a job trace in the Standard Workload Format~\cite{swf} as an input
and then simulates the operation of an HPC system. It can begin with either an idle cluster
or a preloaded cluster and loads jobs sequentially from the job trace. It will query the scheduler
anytime a new job arrives or a job completes and then act accordingly to the returned action.
Additionally, it can perform backfilling when not enough resources are available to schedule the
current job.

\subsection{Metrics for Job Scheduling}
\label{sec:metrics_scheduling}
Job scheduler performance is typically evaluated by a set of optimization goals, commonly known as
scheduling metrics. Each metric reflects specific user requirements and influences the design and
functionality of scheduling approaches. Because no single metric serves as the definitive
standard~\cite{metrics_scheduling}, different strategies emphasize different objectives. Below, we
describe four widely used metrics in job scheduling.

\begin{itemize}
    \item \textbf{Average waiting time:} This metric represents the mean duration, denoted by \( w_j \),
    between a job's submission and the start of its execution.
    
    \item \textbf{Average turnaround time:} Defined as the average interval from the job's submission to
    its completion, this metric is computed as the sum of the waiting time (\( w_j \)) and the job's
    execution time (\( e_j \)):
    \[
    \mathrm{Turnaround\ Time} = w_j + e_j.
    \]

    \item \textbf{Average bounded slowdown:} The traditional slowdown metric is given by
    \[
    \frac{w_j + e_j}{e_j},
    \]
    which can disproportionately penalize short jobs when \( e_j \) is near zero. To mitigate this effect,
    bounded slowdown is defined as
    \[
    \max\left(\frac{w_j + e_j}{\max(e_j, 10)}, 1\right).
    \]
    Here, a minimum execution time of 10 seconds (or another predefined interactive threshold) is enforced,
    ensuring a fairer evaluation across different job durations.
    
    \item \textbf{Resource utilization:} This metric measures the average fraction of compute nodes allocated
    over a given time period, normalized by the total number of nodes in the system:
    \[
    \mathrm{Utilization} = \frac{\mathrm{Allocated\ Nodes}}{\mathrm{Total\ Nodes}} \times 100.
    \]
    It provides a standardized measurement of resource usage.
\end{itemize}

\subsection{HPC Datasets}
\label{sec:dataset}
Two real-world workload traces are used in our study. These traces, summarized in
Table~\ref{tab:traces_overview}, have been merged into a single dataset containing over 11.5 million job
records collected between 2014 and 2020 from two petascale systems—one SGI-8600 cluster and one Dell C6400
chassis cluster. These systems were primarily employed for general-purpose HPC workloads, ranging from
modeling and simulation to visualization.

\begin{table}[h]
\centering
\resizebox{\textwidth}{!}{%
\begin{tabular}{|l|c|c|}
\hline
\textbf{System Name}    & \textbf{Falcon}         & \textbf{Lemhi}          \\ \hline
\textbf{Scheduler}      & PBS                     & PBS                     \\ \hline
\textbf{Compute Nodes}  & 972                     & 504                     \\ \hline
\textbf{Cores}          & 34,992                  & 20,240                  \\ \hline
\textbf{Trace Period}   & 2014/09/24 - 2020/08/04  & 2019/01/23 - 2019/12/31  \\ \hline
\textbf{Number of Jobs} & 11,318,441              & 383,845                 \\ \hline
\textbf{Max Job Length} & 7 days                  & 7 days                  \\ \hline
\end{tabular}%
}
\caption{Overview of Falcon and Lemhi workload traces.}
\label{tab:traces_overview}
\end{table}

\begin{figure}[h]
    \centering
    \begin{subfigure}{0.495\textwidth}
        \centering
        \includegraphics[width=\linewidth]{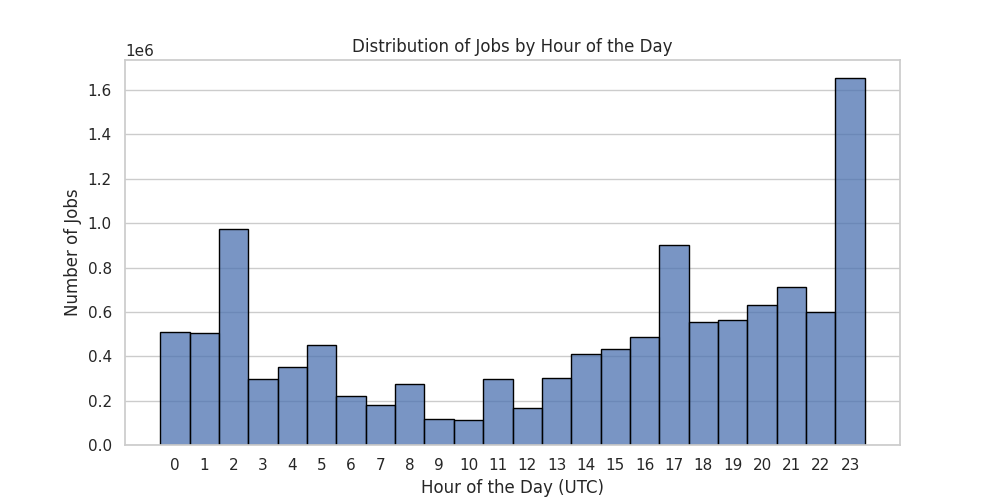}
        \caption{Daily job submission patterns.}
        \label{fig:daily_job_submission_patterns}
    \end{subfigure}
    \hfill
    \begin{subfigure}{0.495\textwidth}
        \centering
        \includegraphics[width=\linewidth]{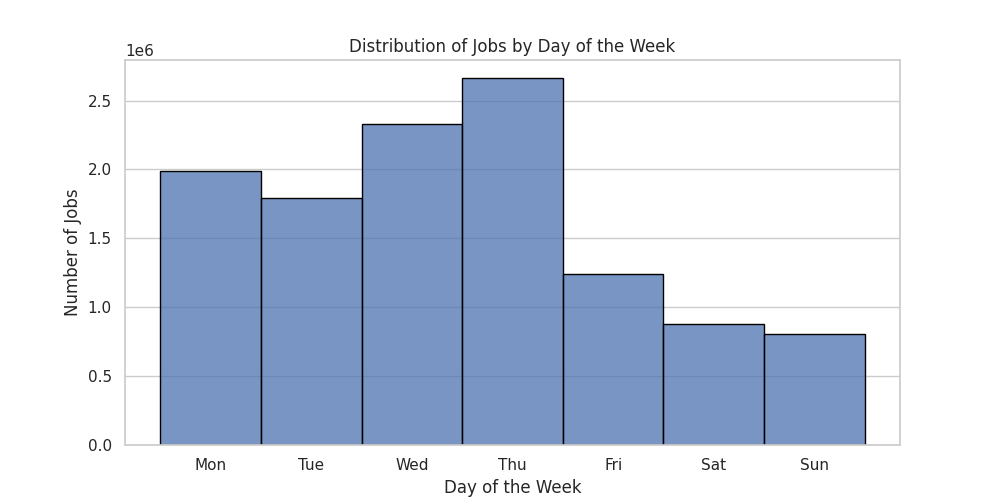}
        \caption{Weekly job submission patterns.}
        \label{fig:weekly_job_submission_patterns}
    \end{subfigure}
    \hfill
    \begin{subfigure}{0.75\textwidth}
        \centering
        \includegraphics[width=\linewidth]{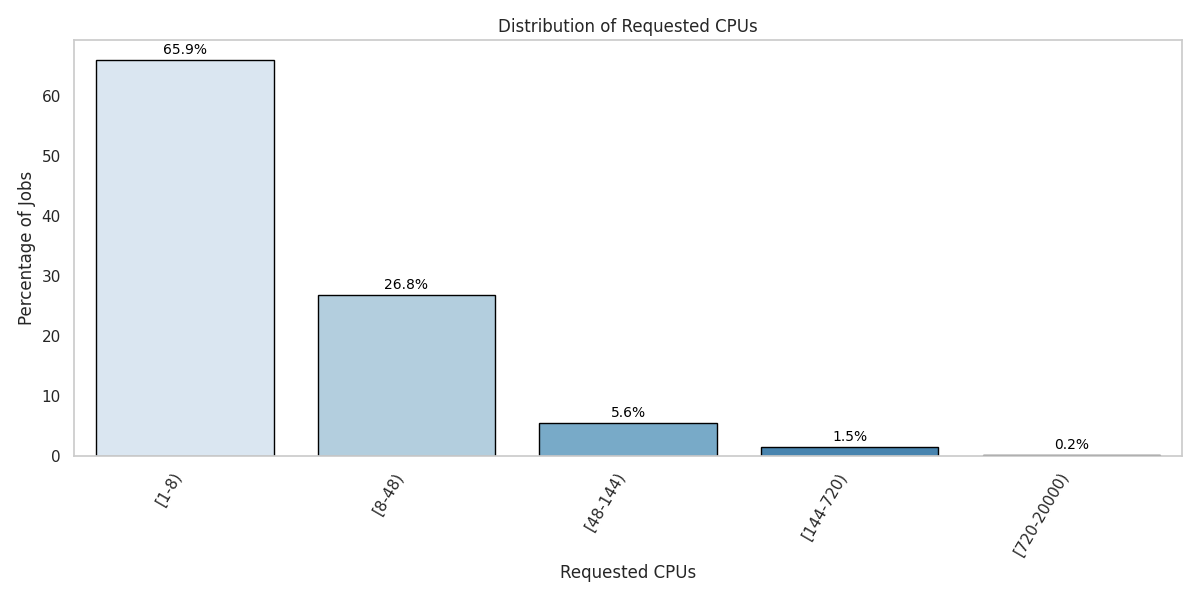}
        \caption{Job size distribution.}
        \label{fig:job_size_distribution}
    \end{subfigure}
    \caption{Characteristics of the Falcon and Lemhi workload traces.}
    \label{fig:traces_distributions}
\end{figure}

These traces were selected because they represent authentic workloads on two general-purpose HPC systems that
support jobs of all sizes. The first workload, coming from the production Falcon system, spans nearly six
years and is derived from a homogeneous system of 972 compute nodes. On Falcon, jobs can range from a single
core up to 34,992 cores. The second workload is taken from the production Lemhi system, a nearly one-year job
log from a homogeneous system of 504 compute nodes that similarly supports jobs ranging from one core to the
full system core count. Figure~\ref{fig:traces_distributions} illustrates key characteristics of the Falcon
and Lemhi traces, such as the distribution of job sizes and the hourly and daily patterns in job submissions.

In addition, the Lublin-256 dataset~\cite{lublin256} is a synthetic workload generated from a well-known
workload model~\cite{lublin}. The synthetic traces simulate realistic job submissions with diverse job sizes,
execution times, and resource demands, thereby capturing a range of scheduling challenges. This dataset
consists of 10,000 job traces and is based on a system with 256 nodes. The SDSC-SP2 HPC job trace
dataset~\cite{sdsc-sp2}, on the other hand, is a real workload record obtained from the IBM SP2 system at the
San Diego Supercomputer Center—a 128-node parallel computing platform. It provides comprehensive scheduling
details, including user accounts, requested versus utilized resources, CPU usage, and timestamps for
submission, waiting, and execution, covering 73,496 jobs collected from May 1998 through April 2000.

The Falcon and Lemhi datasets are used for training our model, while the Lublin-256 and SDSC-SP2 traces serve
exclusively for performance evaluation. All datasets follow the Standard Workload Format (SWF), with each job
attribute treated as a feature. Our model ingests all available job attributes and autonomously determines the
most significant features for effective learning.

\subsection{Implementation and Hyperparameter Tuning}
\label{sec:ray}
Ray~\cite{ray} is a distributed programming framework that enables users to tackle machine learning
tasks at scale. Built atop Ray, RLlib offers a robust library for scalable RL, and
Ray Tune provides advanced hyperparameter search capabilities. Furthermore, Ray’s support for custom Gym
environments allowed us to integrate our job scheduler seamlessly. Leveraging these unified and scalable
libraries, we implemented our RL scheduler using Ray in conjunction with PyTorch, employing the DD-PPO
algorithm along with PBT to boost speed, scalability, and stability.

\begin{figure}[ht]
    \centering
    \includegraphics[width=1.0\linewidth]{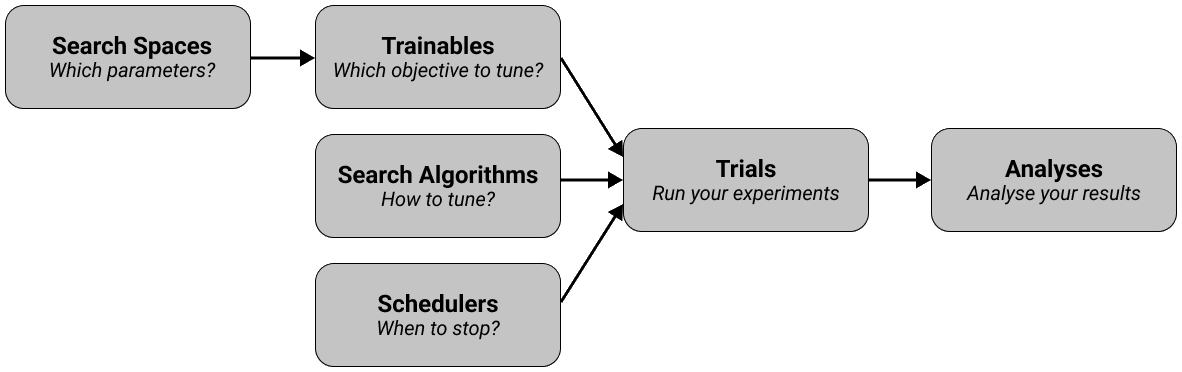}
    \caption{Flow of model training and hyperparameter tuning using the Ray framework.}
    \label{fig:ray_tune_flow}
\end{figure}

Ray Tune integrates effortlessly with RLlib to optimize hyperparameter tuning for RL models. By
leveraging parallel computing and adaptive search algorithms, this collaboration streamlines the
training of RL agents via efficient, episode-based learning. We exploit several features of the Ray
framework, such as PBT, GAE, and automated hyperparameter optimization, to produce the best possible
model. The overall process is illustrated in Figure~\ref{fig:ray_tune_flow} and unfolds as
follows:
\begin{enumerate}
    \item \textbf{Define the RL Environment:} Specify an RL environment for training; in our case, this
    environment is SchedGym.
    \item \textbf{Configure the RL Algorithm:} Employ DD-PPO for fine-tuning within RLlib, leveraging
    Ray Tune for hyperparameter management.
    \item \textbf{Hyperparameter Optimization:} Use Ray Tune to automate the tuning of RL parameters,
    including the learning rate, exploration strategies, and batch size.
    \item \textbf{Train via Episodes:} Allow RL agents to learn through episodes by taking actions,
    receiving rewards, and updating policies.
    \item \textbf{Run Parallel Experiments:} Conduct multiple RL training trials concurrently, across
    either CPU or GPU, to accelerate the search for the optimal configuration.
    \item \textbf{Adaptive Learning:} Utilize PBT for dynamic adjustment of hyperparameters during
    training.
    \item \textbf{Select the Best Model:} Ray Tune evaluates the RL policies based on metrics (in our
    case, \texttt{episode\_reward\_mean}) aligned with our optimization goals, thereby identifying the
    best-performing model.
\end{enumerate}

\section{Results}
\label{sec:results}
The proposed RL-based HPC scheduler, built upon RLScheduler~\cite{rlscheduler}, employs the DD-PPO algorithm
and has been ported to the Ray framework. The scheduler is trained on a large production dataset and evaluated
against various rule-based methods as well as the PPO algorithm. To assess its scheduling performance and
generalization capability, we conducted experiments using four optimization objectives: average waiting time,
average turnaround time, average bounded slowdown, and resource utilization. These metrics are detailed in
Section~\ref{sec:metrics_scheduling}.

\begin{figure}[h]
    \centering
    \begin{subfigure}[b]{0.45\textwidth}
        \centering
        \includegraphics[width=\textwidth]{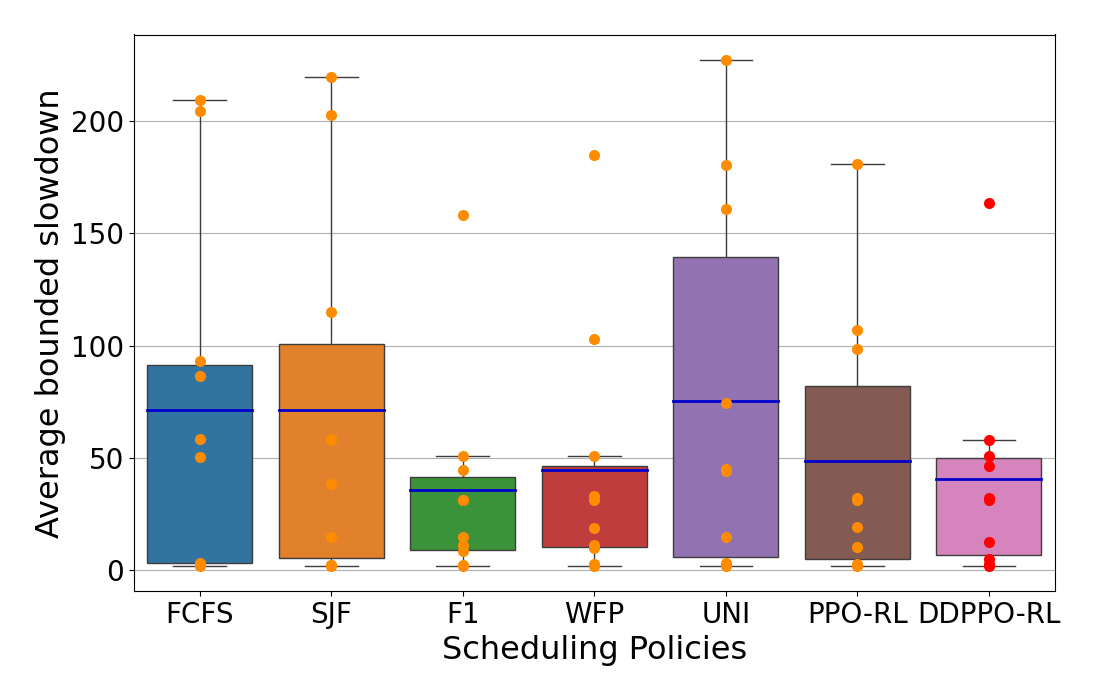}
        \caption{Average bounded slowdown}
        \label{fig:abs_lublin_256}
    \end{subfigure}
    \begin{subfigure}[b]{0.45\textwidth}
        \centering
        \includegraphics[width=\textwidth]{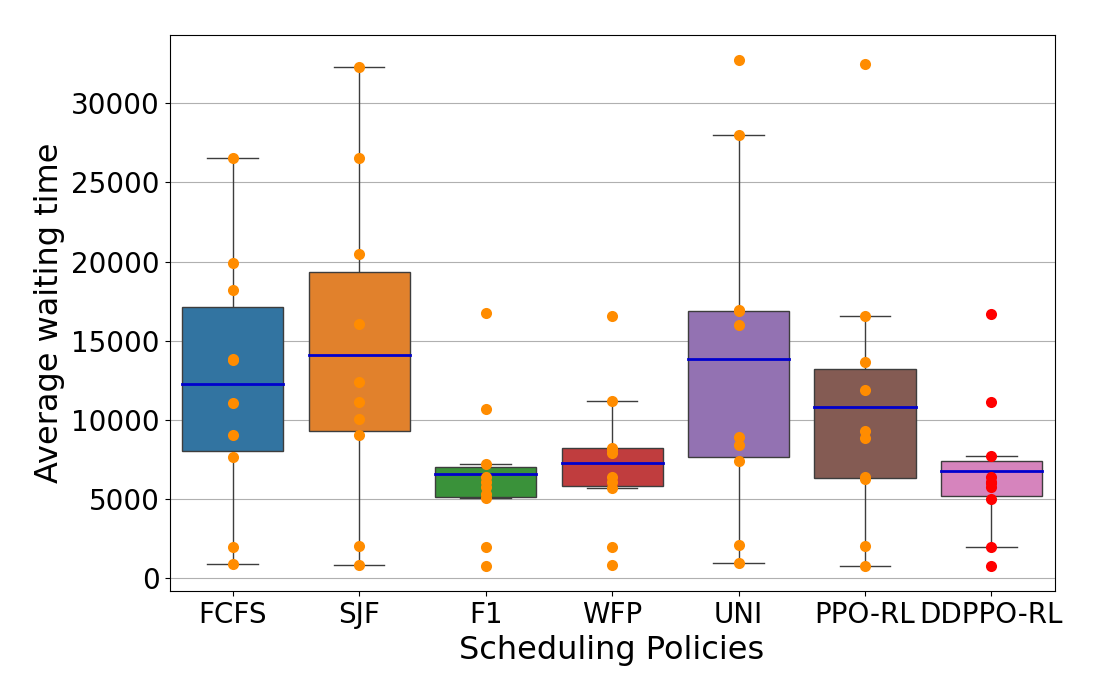}
        \caption{Average waiting time}
        \label{fig:awt_lublin_256}
    \end{subfigure}
    
    \begin{subfigure}[b]{0.45\textwidth}
        \centering
        \includegraphics[width=\textwidth]{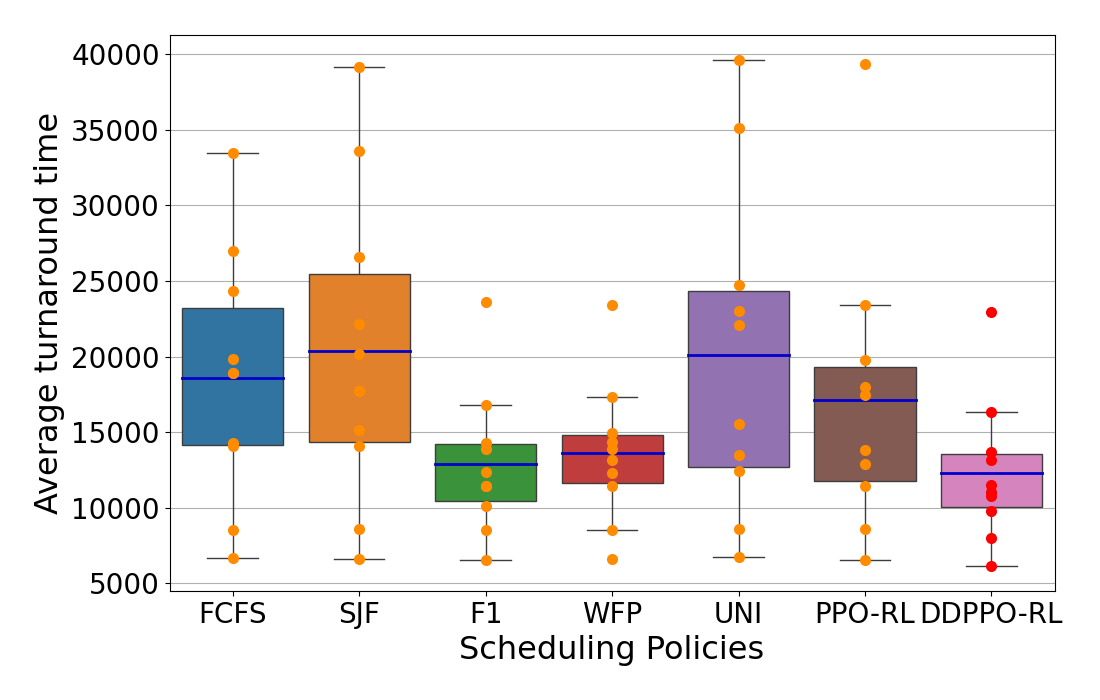}
        \caption{Average turnaround time}
        \label{fig:att_lublin_256}
    \end{subfigure}
    \begin{subfigure}[b]{0.45\textwidth}
        \centering
        \includegraphics[width=\textwidth]{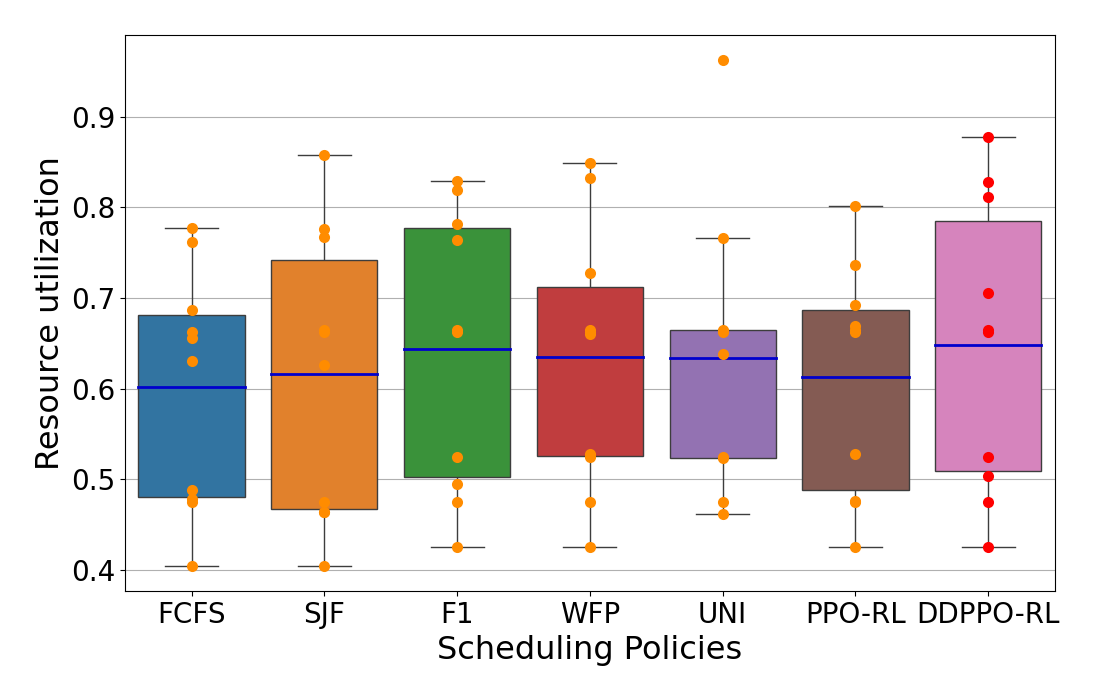}
        \caption{Resource utilization}
        \label{fig:ru_lublin_256}
    \end{subfigure}
    
    \caption{Comparison of the proposed DD-PPO algorithm against several rule-based methods and the PPO
    algorithm using the Lublin-256 dataset (averaged across 10 runs).}
    \label{fig:comparison_lublin_256}
\end{figure}

\begin{figure}[h]
    \centering
    \begin{subfigure}[b]{0.45\textwidth}
        \centering
        \includegraphics[width=\textwidth]{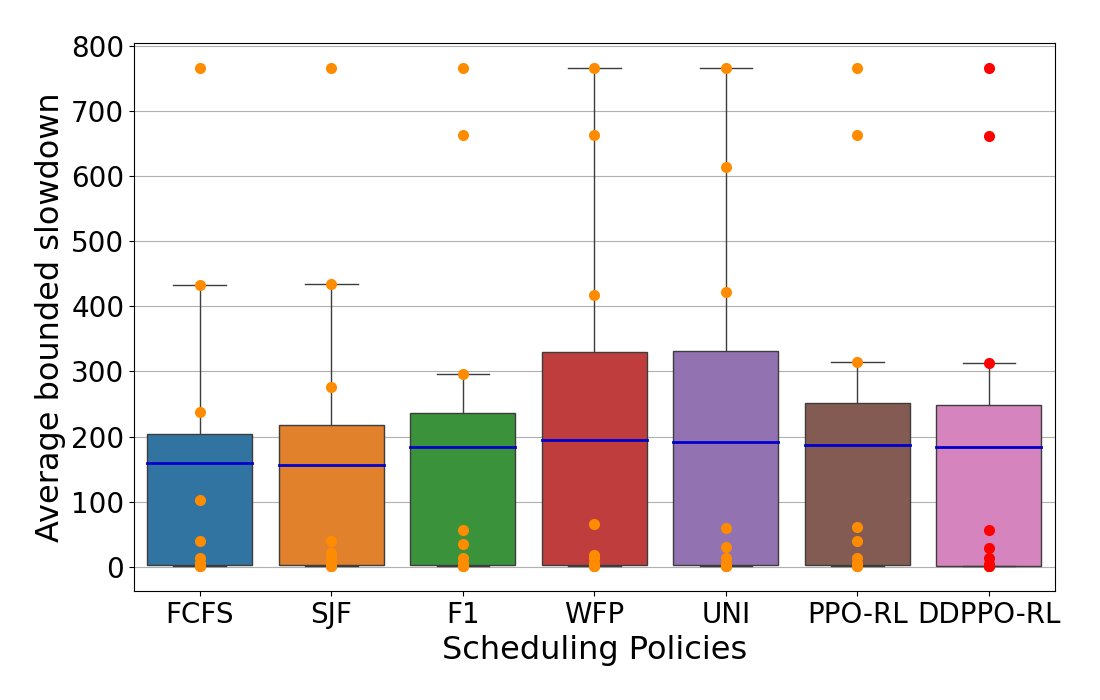}
        \caption{Average bounded slowdown}
        \label{fig:abs_sdsc_sp2}
    \end{subfigure}
    \begin{subfigure}[b]{0.45\textwidth}
        \centering
        \includegraphics[width=\textwidth]{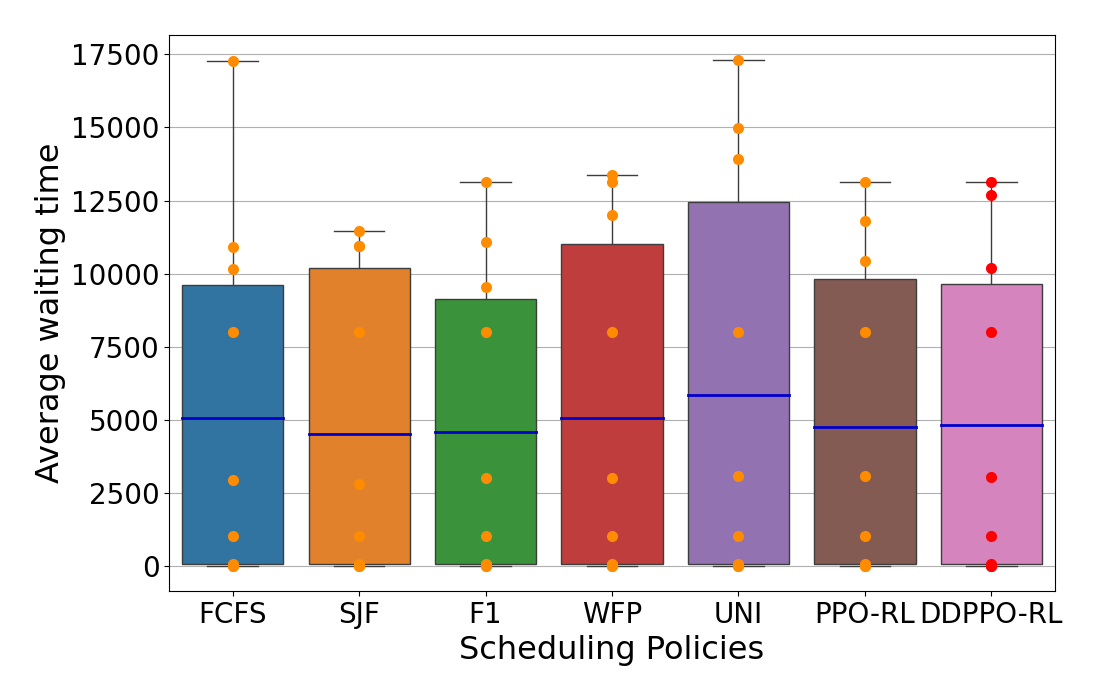}
        \caption{Average waiting time}
        \label{fig:awt_sdsc_sp2}
    \end{subfigure}
    
    \begin{subfigure}[b]{0.45\textwidth}
        \centering
        \includegraphics[width=\textwidth]{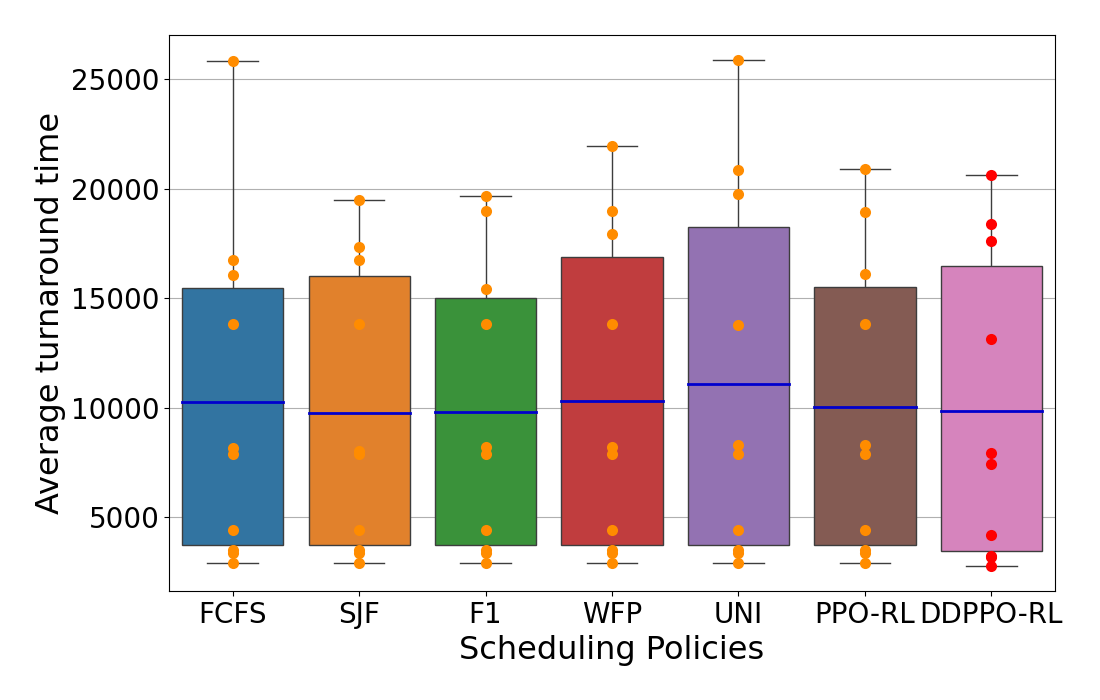}
        \caption{Average turnaround time}
        \label{fig:att_sdsc_sp2}
    \end{subfigure}
    \begin{subfigure}[b]{0.45\textwidth}
        \centering
        \includegraphics[width=\textwidth]{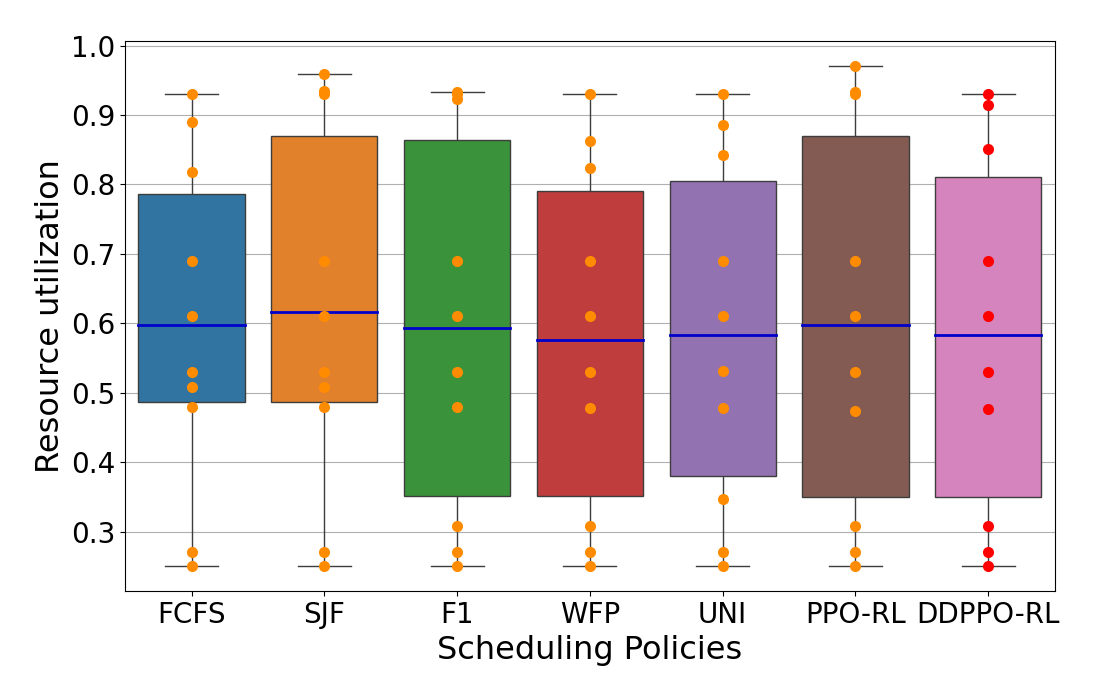}
        \caption{Resource utilization}
        \label{fig:ru_sdsc_sp2}
    \end{subfigure}
    
    \caption{Comparison of the proposed DD-PPO algorithm against several rule-based methods and the PPO
    algorithm using the SDSC-SP2 dataset (averaged across 10 runs).}
    \label{fig:comparison_sdsc_sp2}
\end{figure}

Each scheduling algorithm was tested over 10 iterations, with each iteration processing the same sequence of
1,024 jobs. The resulting box plots, shown in Figures~\ref{fig:comparison_lublin_256}
and~\ref{fig:comparison_sdsc_sp2}, illustrate the scheduling performance: the blue line indicates
the median, the box spans from the $25^{\text{th}}$ to the $75^{\text{th}}$ percentiles, and the orange dots
represent the average values. Notably, the job traces used for these evaluations originate from datasets that
the model had not previously encountered, ensuring an unbiased assessment of its generalization capabilities.

As Figure~\ref{fig:comparison_lublin_256} shows, our method consistently outperforms the PPO
algorithm and most rule-based methods across all optimization objectives. Similarly,
Figure~\ref{fig:comparison_sdsc_sp2} demonstrates that our approach maintains strong performance
across all objectives, with the exception of resource utilization, wherein the PPO algorithm slightly
outperforms our method.

Tables~\ref{tab:mean_SD_lublin-256} and~\ref{tab:mean_SD_sdsc_sp2} report the mean and standard deviation for
each optimization objective for the Lublin-256 and SDSC-SP2 datasets, respectively. These results confirm that
DD-PPO consistently outperforms the PPO algorithm and several rule-based strategies.

\begin{table}[h]
\centering
\resizebox{\textwidth}{!}{%
\begin{tabular}{|l|c|c|c|c|c|c|c|}
\hline
\textbf{Optimization Goal} & \textbf{FCFS} & \textbf{SJF} & \textbf{F1} & \textbf{WFP} & \textbf{UNI} & \textbf{PPO-RL} & \textbf{DDPPO-RL} \\ \hline
Average Bounded Slowdown $\downarrow$   & 71.28 $\pm$ 75.65  & 71.34 $\pm$ 77.69  & 35.47 $\pm$ 44.09  & 44.65 $\pm$ 54.86  & 75.42 $\pm$ 79.42  & 48.64 $\pm$ 57.23  & \textbf{40.37} $\pm$ 45.56  \\ \hline
Average Waiting Time $\downarrow$       & 12293.90 $\pm$ 7559.92  & 14078.40 $\pm$ 9519.50  & 6593.86 $\pm$ 4247.13  & 7287.73 $\pm$ 4221.47  & 13834.10 $\pm$ 9906.34  & 10813.30 $\pm$ 8583.16  & \textbf{6749.03} $\pm$ 4273.09  \\ \hline
Average Turnaround Time $\downarrow$    & 18602.20 $\pm$ 7800.37  & 20386.70 $\pm$ 9866.83  & 12902.20 $\pm$ 4521.77  & 13596.00 $\pm$ 4408.05  & 20142.40 $\pm$ 10361.20  & 17134.90 $\pm$ 8876.82  & \textbf{12322.30} $\pm$ 4411.90  \\ \hline
Resource Utilization $\uparrow$       & 0.60 $\pm$ 0.12  & 0.62 $\pm$ 0.15  & 0.64 $\pm$ 0.15  & 0.63 $\pm$ 0.14  & 0.63 $\pm$ 0.14  & 0.61 $\pm$ 0.12  & \textbf{0.65} $\pm$ 0.15  \\ \hline
\end{tabular}%
}
\caption{Mean $\pm$ standard deviation for various scheduling methods using the Lublin-256 dataset. Bold
values indicate the better performing results between PPO and DDPPO.}
\label{tab:mean_SD_lublin-256}
\end{table}

\begin{table}[h]
\centering
\resizebox{\textwidth}{!}{%
\begin{tabular}{|l|c|c|c|c|c|c|c|}
\hline
\textbf{Optimization Goal} & \textbf{FCFS} & \textbf{SJF} & \textbf{F1} & \textbf{WFP} & \textbf{UNI} & \textbf{PPO-RL} & \textbf{DDPPO-RL} \\ \hline
Average Bounded Slowdown $\downarrow$   & 159.93 $\pm$ 242.2  & 155.79 $\pm$ 247.1  & 183.76 $\pm$ 279.19  & 195.08 $\pm$ 286.96  & 191.24 $\pm$ 279.18  & 186.59 $\pm$ 279.56  & \textbf{184.79} $\pm$ 280.2  \\ \hline
Average Waiting Time $\downarrow$       & 5045.31 $\pm$ 5832.86  & 4526.64 $\pm$ 4883.21  & 4591.25 $\pm$ 4997.6  & 5070.18 $\pm$ 5600.12  & 5842.66 $\pm$ 6714.76  & \textbf{4760.75} $\pm$ 5187.48  & 4822.48 $\pm$ 5289.47  \\ \hline
Average Turnaround Time $\downarrow$    & 10275.4 $\pm$ 7224.19  & 9756.74 $\pm$ 6160.27  & 9821.35 $\pm$ 6259.8  & 10300.3 $\pm$ 6887.67  & 11072.8 $\pm$ 8017.16  & 10021.4 $\pm$ 6520.76  & \textbf{9858.86} $\pm$ 6572.79  \\ \hline
Resource Utilization $\uparrow$       & 0.60 $\pm$ 0.23  & 0.62 $\pm$ 0.25  & 0.59 $\pm$ 0.26  & 0.58 $\pm$ 0.24  & 0.58 $\pm$ 0.24  & 0.60 $\pm$ 0.26  & \textbf{0.62} $\pm$ 0.25  \\ \hline
\end{tabular}%
}
\caption{Mean $\pm$ standard deviation for various scheduling methods using the SDSC-SP2 dataset, with bold
values indicating the best performance between PPO and DDPPO.}
\label{tab:mean_SD_sdsc_sp2}
\end{table}

\begin{figure}[!h]
    \centering
    \begin{subfigure}[b]{0.45\textwidth}
        \centering
        \includegraphics[width=\textwidth]{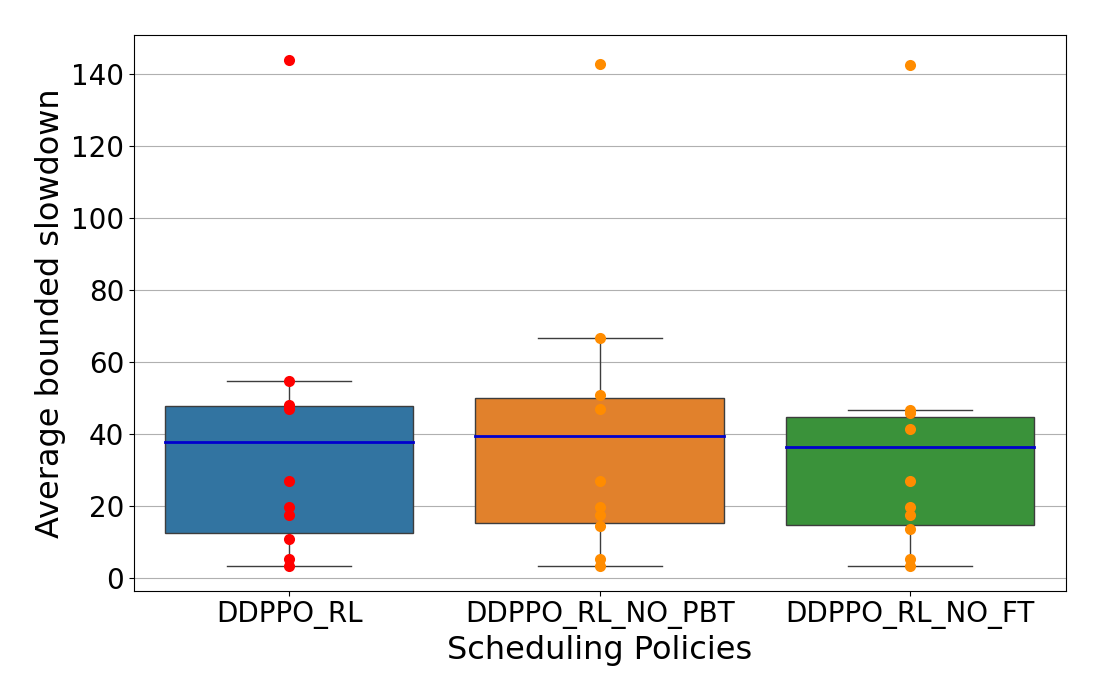}
        \caption{Average bounded slowdown}
        \label{fig:ablation_abs_lublin_256}
    \end{subfigure}
    \begin{subfigure}[b]{0.45\textwidth}
        \centering
        \includegraphics[width=\textwidth]{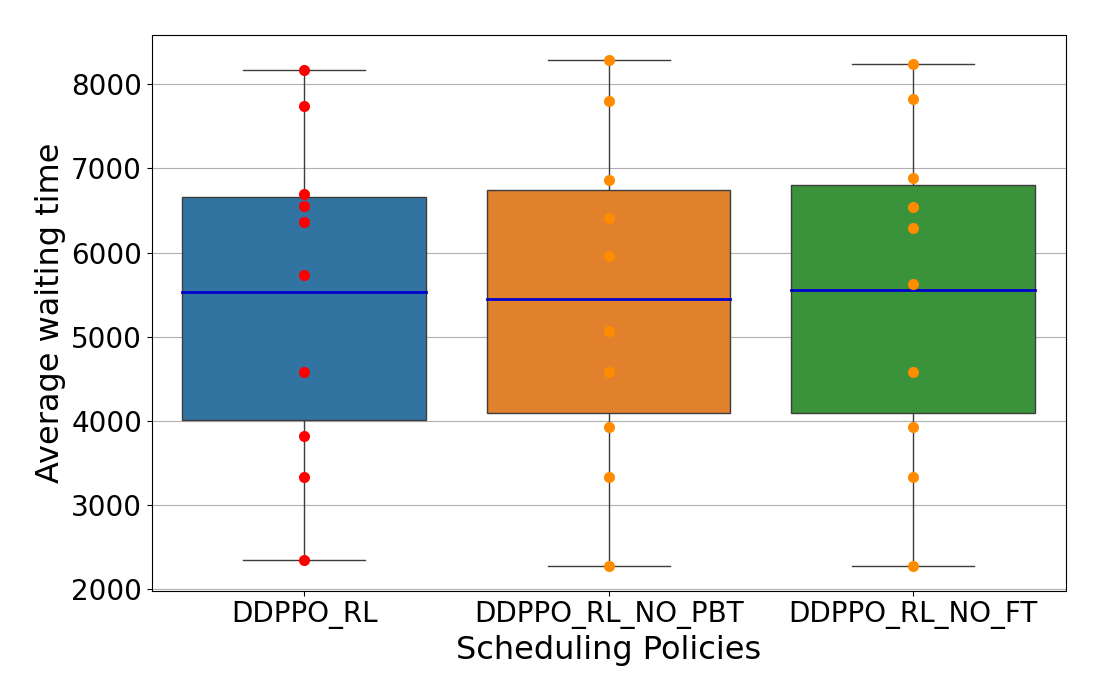}
        \caption{Average waiting time}
        \label{fig:ablation_awt_lublin_256}
    \end{subfigure}
    
    \begin{subfigure}[b]{0.45\textwidth}
        \centering
        \includegraphics[width=\textwidth]{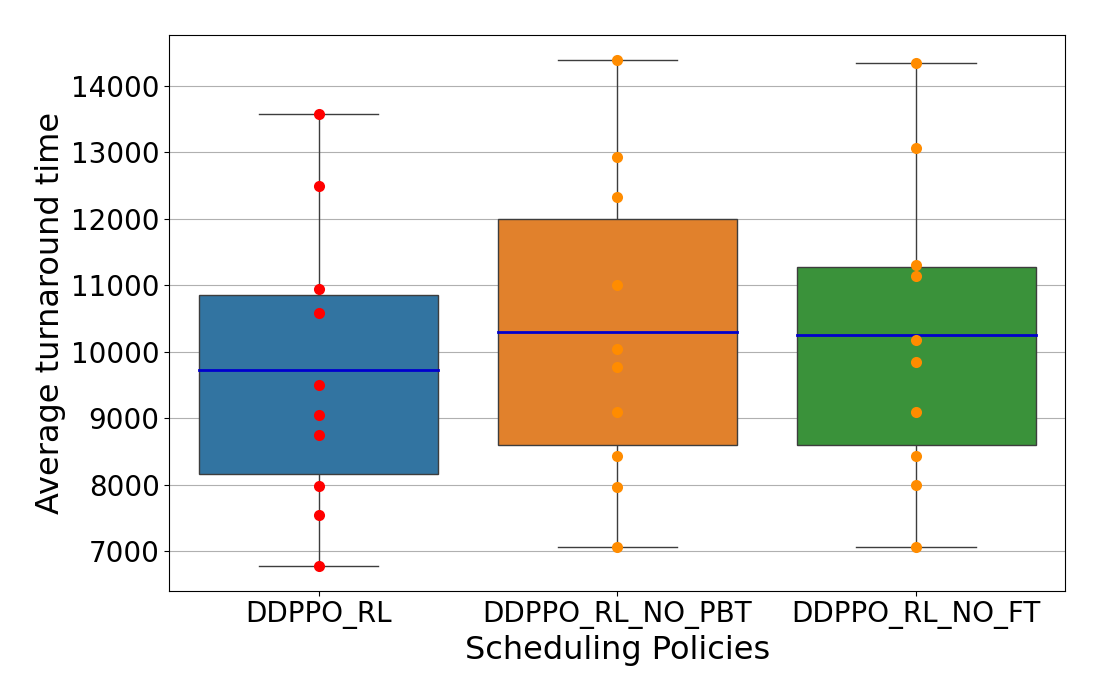}
        \caption{Average turnaround time}
        \label{fig:ablation_att_lublin_256}
    \end{subfigure}
    \begin{subfigure}[b]{0.45\textwidth}
        \centering
        \includegraphics[width=\textwidth]{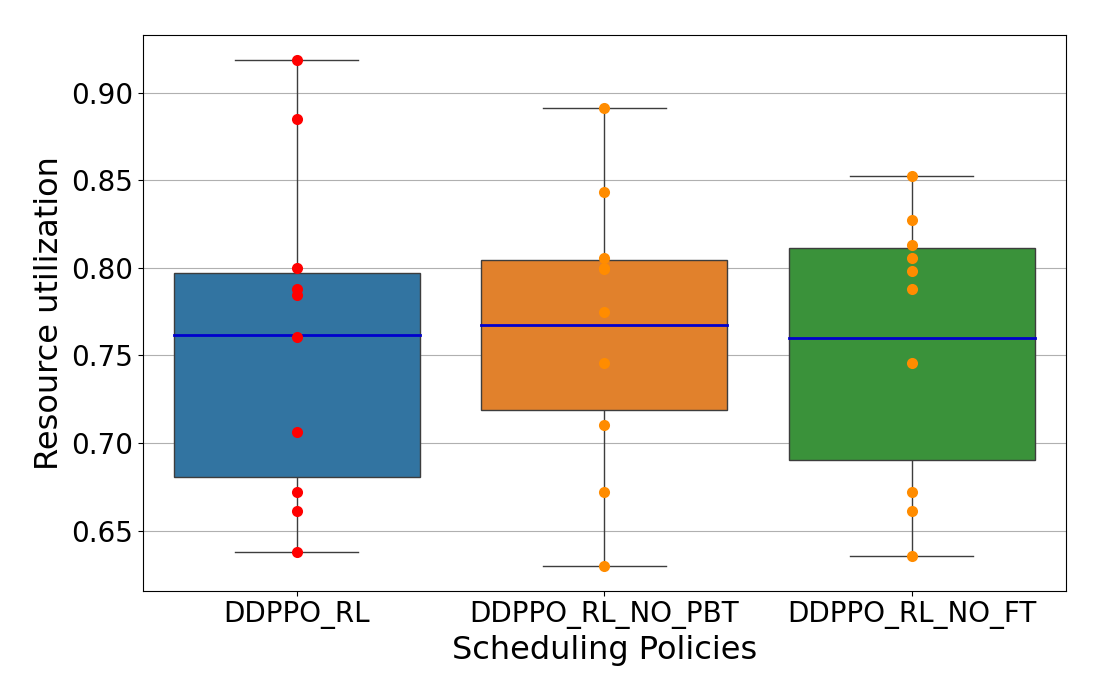}
        \caption{Resource utilization}
        \label{fig:ablation_ru_lublin_256}
    \end{subfigure}
    
    \caption{An ablation study comparison of the proposed DD-PPO algorithm against the algorithm without PBT and without
    FT using the Lublin-256 dataset (averaged across 10 runs).}
    \label{fig:ablation_lublin_256}
\end{figure}

\begin{figure}[!h]
    \centering
    \begin{subfigure}[b]{0.45\textwidth}
        \centering
        \includegraphics[width=\textwidth]{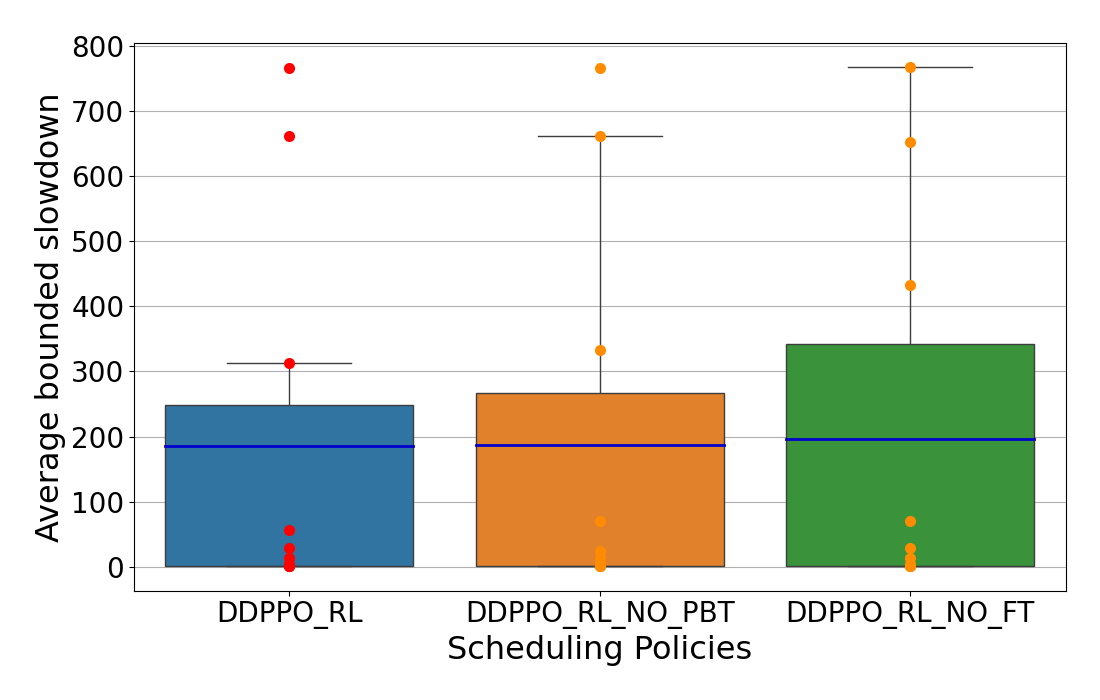}
        \caption{Average bounded slowdown}
        \label{fig:ablation_abs_sdsc_sp2}
    \end{subfigure}
    \begin{subfigure}[b]{0.45\textwidth}
        \centering
        \includegraphics[width=\textwidth]{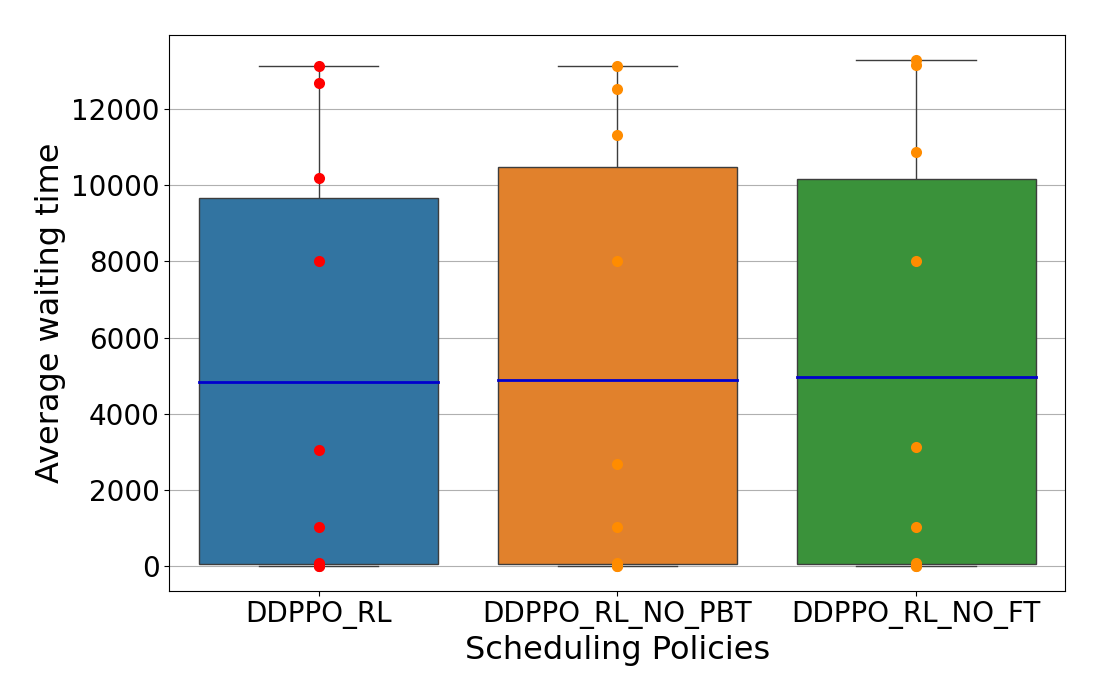}
        \caption{Average waiting time}
        \label{fig:ablation_awt_sdsc_sp2}
    \end{subfigure}
    
    \begin{subfigure}[b]{0.45\textwidth}
        \centering
        \includegraphics[width=\textwidth]{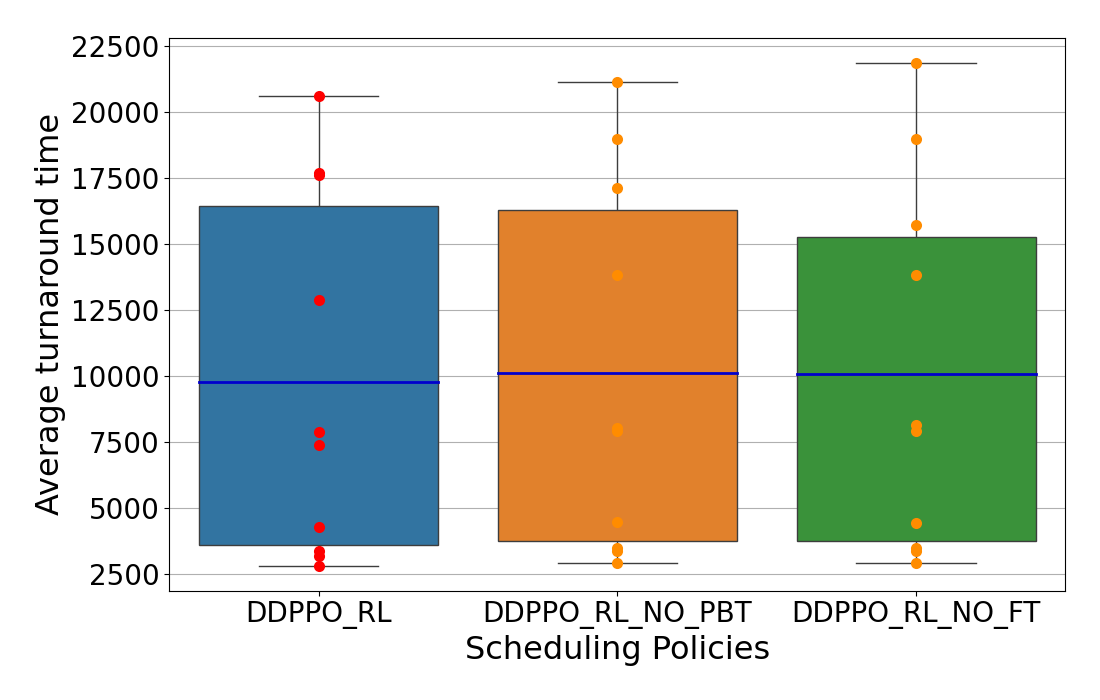}
        \caption{Average turnaround time}
        \label{fig:ablation_att_sdsc_sp2}
    \end{subfigure}
    \begin{subfigure}[b]{0.45\textwidth}
        \centering
        \includegraphics[width=\textwidth]{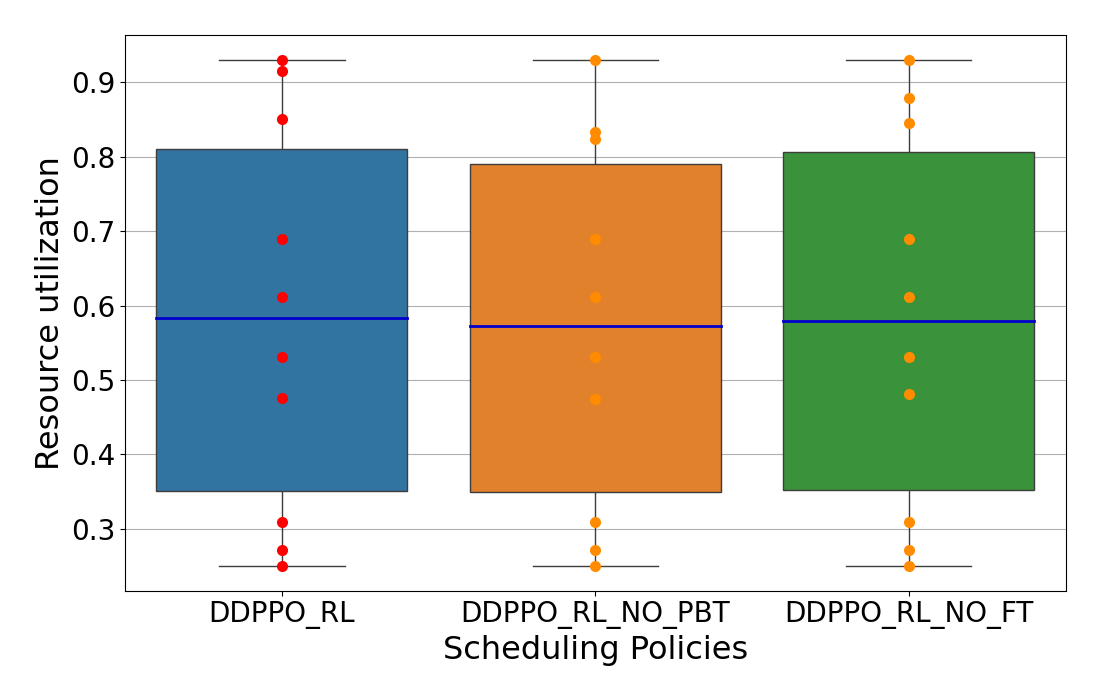}
        \caption{Resource utilization}
        \label{fig:ablation_ru_sdsc_sp2}
    \end{subfigure}
    
    \caption{An ablation study comparison of the proposed DD-PPO algorithm against the algorithm without PBT and without FT
    using the SDSC-SP2 dataset (averaged across 10 runs).}
    \label{fig:ablation_sdsc_sp2}
\end{figure}

To assess the impact of using PBT and Ray’s Hyperparameter Fine-Tuning (FT) during training, an ablation study was conducted. Figures~\ref{fig:ablation_lublin_256} and \ref{fig:ablation_sdsc_sp2} present the results across the Lublin-256 and SDSC SP2
datasets, comparing the fully featured model (with both PBT and FT) against versions where either PBT or FT was removed. The
findings indicate that incorporating PBT and FT consistently improved or maintained performance across all cases for both datasets.

\begin{figure}[!h]
    \centering
    \includegraphics[width=0.7\linewidth]{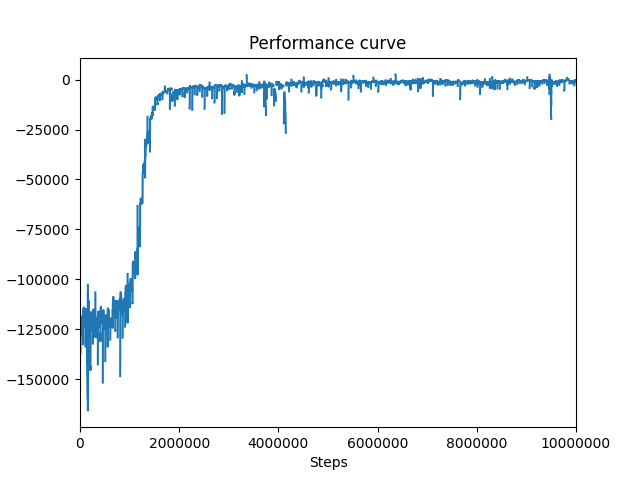}
    \caption{Training reward per iteration of the DD-PPO model.}
    \label{fig:ddppo_reward_performance_curve}
\end{figure}

We selected average bounded slowdown as the optimization goal and designed the reward function accordingly. In
this formulation, the algorithm maximizes its reward by minimizing the average bounded slowdown. The evolution
of the reward function during training is depicted in
Figure~\ref{fig:ddppo_reward_performance_curve}.

The experimental validation demonstrates the advantages of our RL-based approach. First, deep neural networks
effectively capture and represent the system's complex dynamics. Second, the availability of millions of job
records spanning six years furnishes a rich training dataset. Finally, RL allows us to integrate complex,
hard-to-model constraints via an informative reward signal, thereby enhancing scheduling efficiency.

\FloatBarrier
\section{Conclusions}
\label{sec:conclusions}
HPC scheduling is an inherently complex and challenging problem, particularly as HPC systems continue to
grow in complexity. RL offers a viable approach to addressing this challenge. However, RL models typically
require large volumes of data to develop an effective generalization capability.

To tackle this issue, our study focuses on optimizing HPC scheduling by leveraging the DD-PPO algorithm,
trained on extensive real-world workload data. This approach enhances model robustness by incorporating
diverse data collected from multiple workers across a large dataset, improving generalization and adaptability
to previously unseen scenarios.

Experimental results confirm the superior robustness of our model compared to
rule-based algorithms and the PPO algorithm. Through evaluations on previously unseen HPC job traces, we
demonstrate that our model consistently surpasses PPO across four key optimization objectives: Average waiting
time, Average turnaround time, Average bounded slowdown, and Resource utilization. These findings affirm that
combining DD-PPO with large-scale datasets results in a more generalized and effective model than a standard
PPO-based approach trained on the same dataset. Finally, future work would encompass utilizing our large dataset
on other RL HPC scheduling algorithms to quantify the impact on their performance.

\printbibliography

\end{document}